\documentclass[pre,twocolumn,showpacs,preprintnumbers,amsmath,amssymb]{revtex4}

\usepackage{graphicx}
\usepackage{dcolumn}
\usepackage{bm}

\begin{document}


\title{Dependence on dilution of critical and compensation temperatures
of a two dimensional mixed spin-1/2 and spin-1 system}

\author{Ekrem Aydiner} \email{ekrem.aydiner@deu.edu.tr}
\author{Yusuf Y\"{u}ksel}
\author{Ebru K{\i}\d{s}-\c{C}am}
\author{Hamza Polat}

\affiliation{ Department of Physics, Dokuz Eyl\"{u}l University,
TR-35160 Izmir, Turkey}
\date{\today}

\begin{abstract}
In this study, dependence on site dilution of critical and
compensation temperatures of a two dimensional mixed spin-1/2 and
spin-1 system has been investigated with Monte Carlo simulation. The
dependence of the thermal and magnetic behaviors on dilution of
mixed spin system have been discussed. We have shown that the
critical and compensation temperatures of diluted mixed spin system
linearly decrease with increasing number of diluted sites.
\end{abstract}

\maketitle


Ferrimagnetic systems i.e., materials have got considerable
attention due to their technological applications. These systems
consist of different spin sublattices and, therefore, are so-called
as mixed spin systems. One of the most important properties of these
systems is that they have a compensation temperature. In a
ferrimagnetic spin system, sublattices have inequivalent moments
interacting antiferromagnetically. At low temperatures, even though
inequivalent moments of sublattices are antiparallel, however, they
may not cancel each other owing to different temperature
dependencies of the sublattice magnetization. Fortunately,
sublattices of ferrimagnetic system compensate each other completely
at $T=T_{comp}$ below the N\'{e}el temperature \cite{Neel}. This
critical point value is called as compensation temperature or
compensation point of ferrimagnetic spin system. At the compensation
point, the total magnetization of ferrimagnetic system vanishes and
only a small driving field is required to reverse the sign of
magnetization of a locally heated magnetic domain by using a focused
laser beam. Hence, writing and erasing processes can be achieved at
this point. This kind of ferrimagnetic systems are known as
magneto-optic materials.

In technological applications, it is important to produce the
material i.e., spin system which has a desired compensation
temperature. In recent studies, it has been shown that dilution
plays a role on the compensation point of ferrimagnetic mixed spin
system. Therefore, the effect of dilution on the compensation
temperature of mixed spin systems have been studied and interesting
phenomenon have been found. For example, Bob\'{a}k and
Ja\v{s}\v{c}ur \cite{Bobak1} investigated the magnetization and
phase diagrams of a diluted honeycomb lattice which consists of
spin-1/2 and spin-1 sub-lattices within the framework of an
effective-field theory with correlations by neglecting the crystal
field interaction. They obtained a number of interesting phenomena,
such as possibility of two compensation points in the total
magnetization curve. By using an effective-field theory with
correlations, Xin et al., \cite{Xin1} investigated the ground state
and finite temperature properties of a mixed spin-1/2 and spin-1
ferrimagnetic system on square and honeycomb lattices in which the
magnetic atoms are randomly diluted with non-magnetic atoms by
neglecting the crystal field interaction. They found that the
transition temperature of the square lattice is higher than that of
the honeycomb lattice and the compensation range of the square
lattice is larger than that of the honeycomb lattice with the same
parameters. They also observed two compensation points on square
lattice. Xin et al., \cite{Xin2} also discussed the influence of the
crystal field interaction and concentration of the non-magnetic
atoms on the magnetic and thermal properties of mixed spin-1/2 and
spin-1 honeycomb and square lattices. They observed the tricritical
point, reentrant phenomena and two compensation points on square
lattice, and also observed reentrant phenomena and two compensation
points on honeycomb lattice. Bob\'{a}k and Jur\v{c}i\v{s}in
\cite{Bobak2} investigated the critical and compensation
temperatures and total magnetization of a diluted mixed spin-1 and
spin-3/2 Ising ferrimagnetic system by using an effective-field
approximation. They showed that the system exhibits three
compensation points in certain concentration ranges of magnetic
atoms. Kaneyoshi et al. \cite{Kaneyoshi}, discussed magnetic
properties of a diluted spin-2 and spin-5/2 ferrimagnetic honeycomb
lattice on basis of effective-field theory. They found that two or
three compensation points are possible in diluted system with
special values of crystal field and concentration of magnetic atoms.
In addition, the influence of a transverse field on the compensation
point were examined on honeycomb and square lattices
\cite{Nakamura}.

As it can be seen from previous studies dilution effects on the
thermal and magnetic behavior of the ferrimagnetic mixed systems
have been investigated in detail. However, dependence of the
critical and compensation temperatures on dilution has never been
taken into account up to now. Therefore, this point still deserves
particular attention for mixed spin systems. In this paper, we
discuss dependence of the critical and compensation temperatures on
dilution in a two dimensional mixed spin-1/2 and spin-1 system with
Monte Carlo simulation method. We find that the critical and
compensation temperatures linearly decrease as the number of diluted
sites increases in two dimensional mixed spin-1/2 and spin-1 system.


In order to investigate the effect of dilution on the critical and
compensation temperatures and on other physical quantities, we
consider antiferromagnetically interacting two dimensional mixed
spin ($1/2$,$1$) system with Hamiltonian
\begin{equation}\label{1}
H=-J_{1}\sum_{<nn>}\sigma_{i}S_{j}-J_{4}\sum_{<nnn>}\sigma_{i}\sigma_{k}+D\sum_{j}S_{j}^{2}
\end{equation}
where $\sigma=\pm1/2$ and $S=\pm1,0$. The first sum in Eq.\,(1) is
over the nearest-neighbor and the second one is over the
next-nearest-neighbor spins. $J_{1}$ ($J_{1}<0$) and $J_{4}$
parameters define exchange interactions between the neighbor spins,
and $D$ is the crystal field.


To simulate the system, we initially chose spins $S$ and $\sigma$
randomly, but with equal number, in a discrete lattice according to
Eq.\,(1). We also set equal number of non-magnetic atoms randomly in
each sub-lattice. Then, we employed Metropolis Monte Carlo
simulation algorithm \cite{Binder} to the Eq.\,(1) on a $L\times L$
square lattice with periodic boundary conditions. Configurations
were generated by selecting the sites in sequence through the
lattice and making single-spin-flip attempts, which were accepted or
rejected according to the Metropolis algorithm. Data was generated
over $100$ realization for $L=16,32$ and $64$ with different numbers
of non-magnetic atoms by using $25000$ Monte Carlo steps per site
after discarding the first $2500$ steps. Our program calculates the
sub-lattice magnetizations $M_{A}$ and $M_{B}$, the total
magnetization $M$, the magnetic susceptibility $\chi$ and the
specific heat $C$ for different densities of diluted sites. These
quantities are defined as
\begin{equation}
M_{A}=\frac{2}{L^{2}} \left\langle\sum_{j}S_{j}\right\rangle
\end{equation}
\begin{equation}
M_{B}=\frac{2}{L^{2}} \left\langle\sum_{i}\sigma_{i}\right\rangle
\end{equation}
\begin{equation}
M=\frac{1}{2}\left(M_{A}+M_{B}\right)
\end{equation}
\begin{equation}
\chi=\frac{1}{kT}\left(\langle M^{2}\rangle-\langle
M\rangle^{2}\right)
\end{equation}
\begin{equation}
C=\frac{1}{kT^{2}}\left(\langle E^{2}\rangle-\langle
E\rangle^{2}\right)
\end{equation}
where $T$ denotes temperature, $E$ is the internal energy of the
system, and $k$ is Boltzmann constant (here $k=1$).

To determine the compensation temperature $T_{comp}$ from the
computed magnetization data, the intersection point of the
absolute values of the sub-lattice magnetizations was found using
the relations
\begin{equation}
\left|M_{A}(T_{comp})\right|=\left|M_{B}(T_{comp})\right|
\end{equation}
\begin{equation}
sign(M_{A}(T_{comp}))=-sign(M_{B}(T_{comp}))
\end{equation}
with $T_{comp}<T_{c}$, where $T_{c}$ is the critical i.e., N\'{e}el
temperature. Eqs.\,(7) and (8) indicates that the sign of the
sub-lattice magnetizations are different, however, absolute values
of them are equal to each other at the compensation point.


In order to see the effect of the non-magnetic atoms on the magnetic
and thermal behaviors of two dimensional mixed spin system defined
with Eq.\,(1), choosing $J_{1}=-2$, $J_{4}=8$ and $D=2.6$ we give
the results of $64\times64$ square lattice for $N=0$, $64$, $256$,
$512$ in Figs.\,1-6. Here we must state that the nearest-neighbor
interaction $J_{1}$ in the Eq.\,(1) does not play a role on the
compensation temperature, whereas observation of the compensation
temperature depends on the parameters $J_{4}$ and $D$. This point
has been discussed in Ref.\,\cite{Buendia}. Therefore, in this
study, we have particularly focused on the effect of the different
dilution rates with non-magnetic atoms $N$ for arbitrary fixed
$J_{1}$, $J_{4}$ and $D$. On the other hand, to analyze the effect
of the non-magnetic atoms on the critical and compensation
temperatures, for fixed $J_{1}=-2$, $J_{4}=8$ and two different
crystal field values $D=1.6$ and $D=2.6$ we give results of
different lattice size ($L=16$, $32$, $64$) for different densities
of diluted site in Figs.\,7 and 8.

\begin{figure}
\includegraphics[width=9cm, height=8cm]{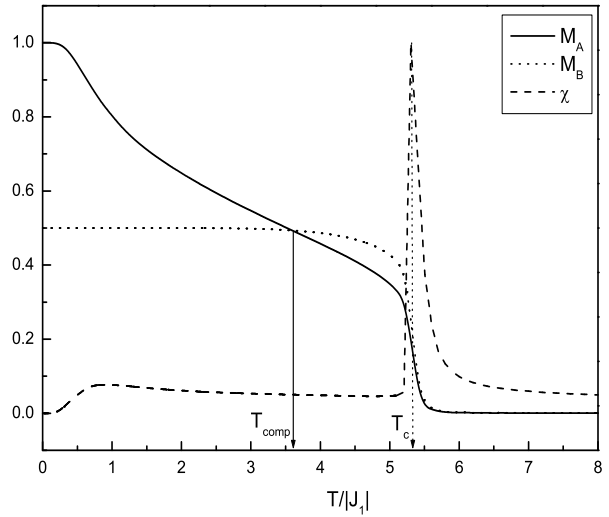}
\caption{\label{fig:epsart} The temperature dependencies of
sublattice magnetizations $M_{A}$, $M_{B}$ and susceptibility $\chi$
of non diluted square lattice for $J_{1}=-2$, $J_{4}=8$ and
$D=2.6$.}
\end{figure}
\begin{figure}
\includegraphics[width=9cm, height=8cm]{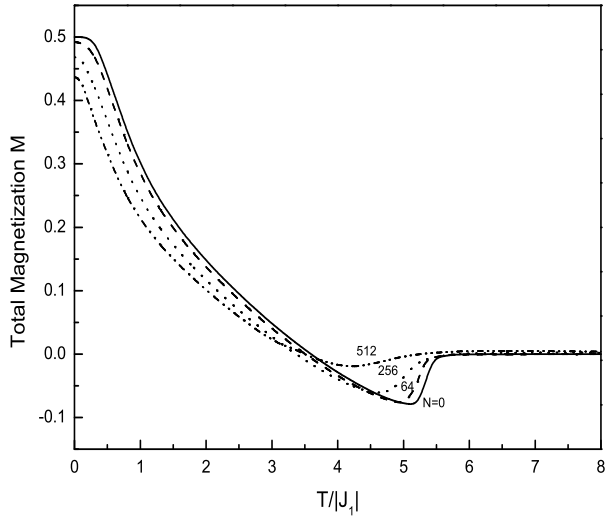}
\caption{\label{fig:epsart} Total magnetization $M$ versus reduced
temperature $T/|J_{1}|$ for different $N$ with $J_{1}=-2$, $J_{4}=8$
and $D=2.6$.}
\end{figure}
\begin{figure}
\includegraphics[width=9cm, height=8cm]{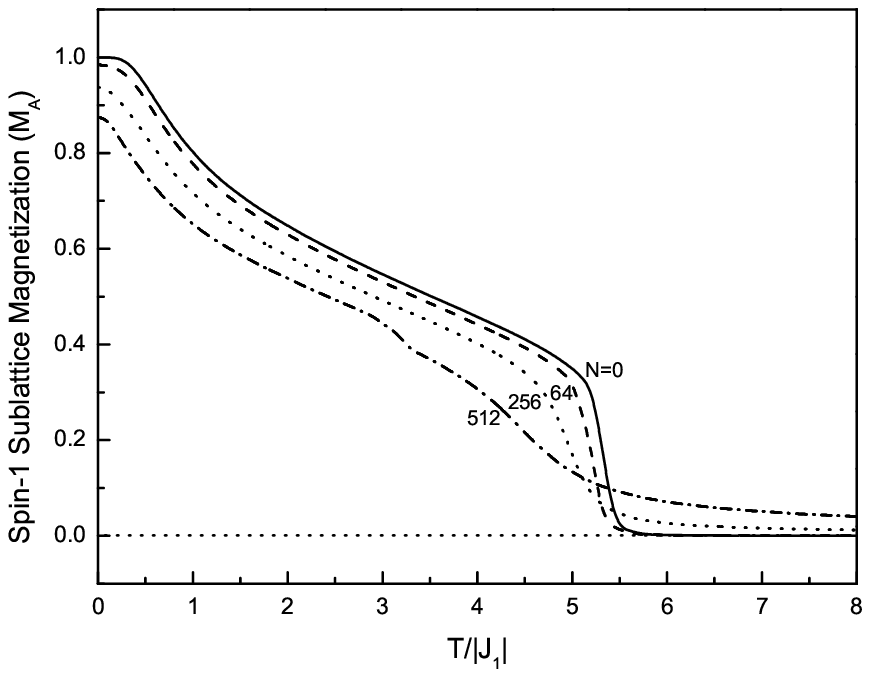}
\caption{\label{fig:epsart} Spin-1 sublattice magnetization $M_{A}$
versus reduced temperature $T/|J_{1}|$ for different $N$ with
$J_{1}=-2$, $J_{4}=8$ and $D=2.6$.}
\end{figure}
\begin{figure}
\includegraphics[width=9cm, height=8cm]{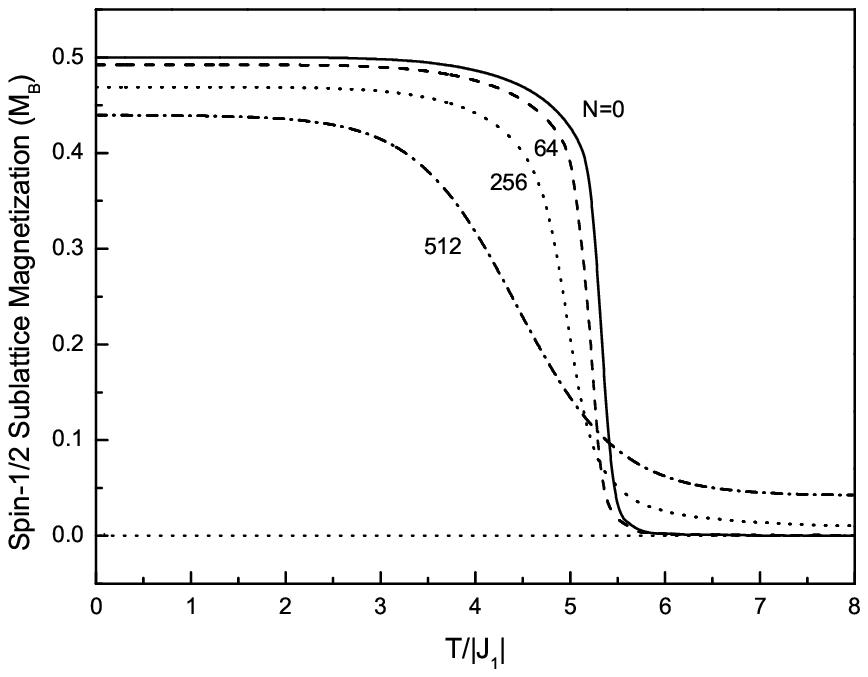}
\caption{\label{fig:epsart} Spin-1/2 sublattice magnetization
$M_{B}$ versus reduced temperature $T/|J_{1}|$ for different $N$
with $J_{1}=-2$, $J_{4}=8$ and $D=2.6$.}
\end{figure}
\begin{figure}
\includegraphics[width=9cm, height=8cm]{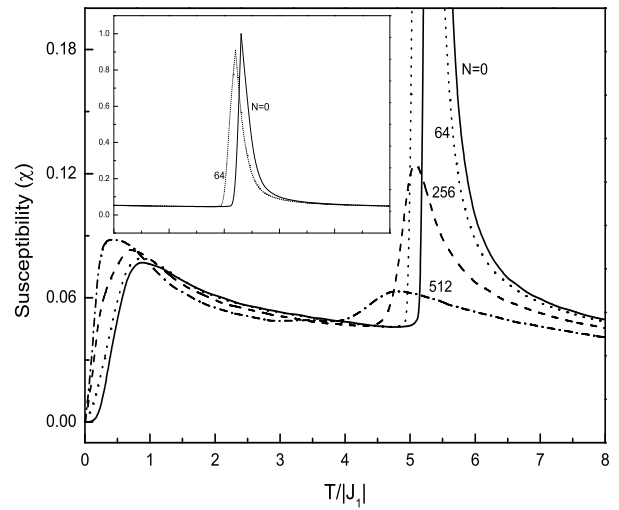}
\caption{\label{fig:epsart} Magnetic susceptibility $\chi$ versus
reduced temperature $T/|J_{1}|$ for different $N$ with $J_{1}=-2$,
$J_{4}=8$ and $D=2.6$. }
\end{figure}


In Fig.\,1, temperature dependencies of sublattice magnetizations
$M_{A}$, $M_{B}$ and susceptibility $\chi$ of non-diluted square
lattice are given for $J_{1}=-2$, $J_{4}=8$ and $D=2.6$. As
expected, for chosen parameters the mixed spin system has a
compensation point near $T/|J_{1}|=3.6$ and a critical point near
$T/|J_{1}|=5.3$ as seen from Fig.\,1. The results of $M_{A}$,
$M_{B}$ and $\chi$ for these parameters are in an excellent
agreement with the results of Ref.\,\cite{Buendia}.


In Fig.\,2, total magnetization versus temperature has been plotted
for several numbers of non-magnetic atoms $N$ ($N=0$, $64$, $256$,
$512$) for fixed $J_{1}=-2$, $J_{4}=8$ and $D=2.6$. In this figure,
there are two zeros of magnetization curves for different number of
non-magnetic atoms. The first zero indicates that the temperature
value at which total magnetization $M$ is zero which corresponds to
the compensation temperature point, and on the other hand, the
second zero denotes that the temperature value at which $M$ is zero
which corresponds to critical temperature point. Though one can see
some quantitative differences in the total magnetization curves for
different number of non-magnetic atoms, but it is hard to comment on
this naive picture about dependence of dilution of the critical and
compensation temperatures. However, it might be also predicted the
compensation and critical points appear near $T/|J_{1}|=3.6$ and
$T/|J_{1}|=5.3$ for different number non-magnetic atoms,
respectively.
\begin{figure}
\includegraphics[width=9cm, height=8cm]{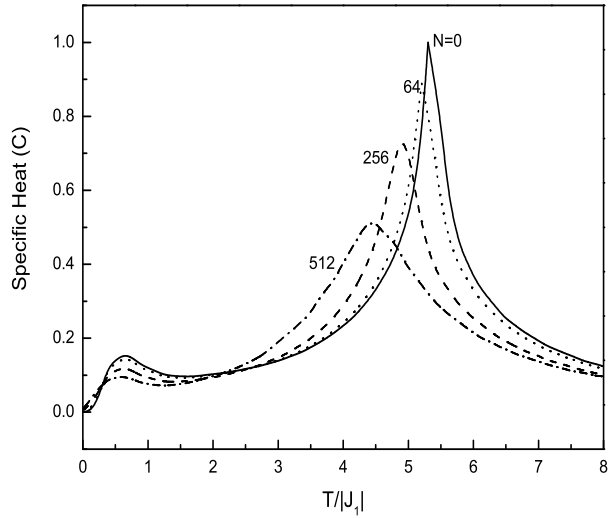}
\caption{\label{fig:epsart} Temperature dependence of the specific
heat $C$ for different $N$ values with $J_{1}=-2$, $J_{4}=8$ and
$D=2.6$.}
\end{figure}


In Figs.\,3 and 4, for different $N$ values ($N=0$, $64$, $256$,
$512$) and fixed $J_{1}=-2$, $J_{4}=8$ and $D=2.6$ values, the
behavior of spin-1 and spin-1/2 sublattice magnetizations $M_{A}$
and $M_{B}$ have been separately demonstrated, respectively. As seen
from these figures, sublattice magnetizations take values which are
different from zero at the compensation point while, as seen in
Fig.\,2, the total magnetization value is zero at the compensation
point for all values of $N$. These two figures inform us about
behavior of compensation points for different $N$ values. Indeed, it
is possible to obtain the compensation points for different $N$
values if $M_{A}$ is mapped onto $M_{B}$ in the same figure as well
as in Fig.\,1. The first crosses in the sublattice magnetization
curves of $M_{A}$ and $M_{B}$ correspond to compensation points. One
can easily predict from these figures that the compensation
temperatures decrease as the number of non-magnetic atoms $N$
increases. The dilution dependence of compensation temperature which
we have obtained from Figs.\,3 and 4 is given below. However, it is
still hard to say from these figures about the dependence of
dilution of critical temperature points.
But it is possible to see the behavior of the critical temperature
points for different $N$ values in Fig.\,5 in which magnetic
susceptibility has been plotted as a function of temperature for the
same values of parameters. As it can be seen from Fig.\,5 that there
are two relatively sharp peaks and humps in magnetic susceptibility
curves for different $N$ values. For the chosen parameters, the
critical temperature points can be easily determined from the second
peaks and humps. Because, the second peaks and humps in the
susceptibility curves appear at critical temperatures and indicate a
phase transition from ferrimagnetic to paramagnetic, as expected. It
can be clearly seen that the maximum points of the susceptibility
curves become smaller and slide to left when the values of $N$
increase, which means that critical temperature decreases and
magnetization gets weaker with increasing $N$. The dilution
dependence of critical temperature which we have obtained from
Fig.\,5 is given below. On the other hand, the first humps in
Fig.\,5 may probably originate from the crystal field $D$, and they
do not indicate a compensation temperature or a phase transition in
the model, however, first humps in the susceptibility curves are
also affected by the existence of the non-magnetic atoms in the
lattice.


In Fig.\,6, temperature dependence of the specific heat has been
plotted for different $N$ and fixed  $J_{1}=-2$, $J_{4}=8$, $D=2.6$
values. Similarly, there are also two relatively sharp humps and
peaks in specific heat. The first humps probably may also originate
from $D$, as well as the first humps in magnetic susceptibility. The
second humps and peaks denote a phase transition. As seen from
Fig.\,6, maxima of the specific heat curves become smaller and slide
to left when the value of $N$ increases. It confirms that the
critical temperature decreases with increasing $N$.

\begin{figure}
\includegraphics[width=9cm, height=8cm]{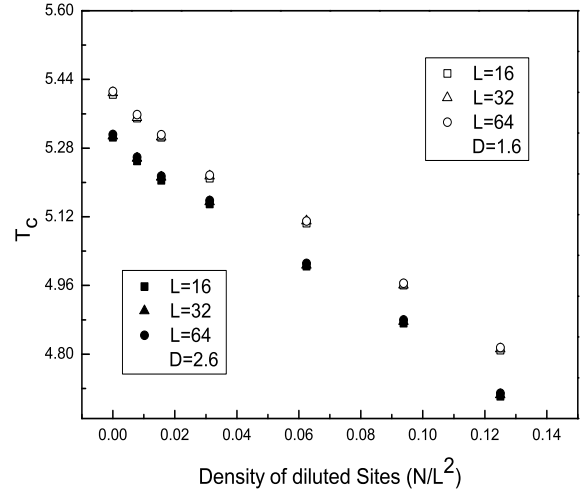}
\caption{\label{fig:epsart} Dependence on the density of diluted
site of the critical temperature for fixed $J_{1}=-2$, $J_{4}=8$ and
two different crystal field values $D=1.6$ and $D=2.6$.}
\end{figure}
Fig.\,7 shows the dependence on the density of diluted site of the
critical temperature for fixed $J_{1}=-2$, $J_{4}=8$ and two
different crystal field values $D=1.6$ and $D=2.6$ with different
lattice sizes. As it can be seen from figure, non-magnetic atoms
play an important role on the critical temperature of the system.
Indeed, critical temperature of diluted system linearly decreases
for fixed $J$ and $D$ with increasing density of diluted site in the
lattice. Similarly crystal field also affects the critical
temperature of the system. It can be seen from Fig.\,7 critical
temperature of the system systematically decreases when the crystal
field value is increased. On the other hand, we can say that the
critical temperature of the diluted mixed system does not affected
from the lattice size.

\begin{figure}
\includegraphics[width=9cm, height=8cm]{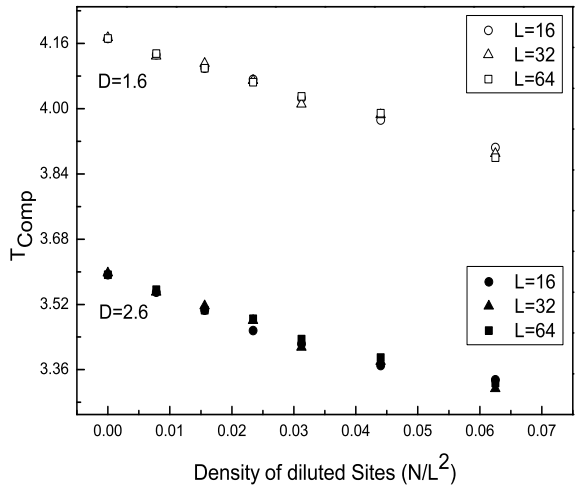}
\caption{\label{fig:epsart} Dependence on the density of diluted
site of the compensation temperature for fixed $J_{1}=-2$, $J_{4}=8$
and two different crystal field values $D=1.6$ and $D=2.6$.}
\end{figure}
Fig.\,8 shows the dependence on the density of diluted site of the
compensation temperature for fixed $J_{1}=-2$, $J_{4}=8$ and two
different crystal field values $D=1.6$ and $D=2.6$ with different
lattice sizes. Similarly, non-magnetic atoms and crystal field play
a significant role on the compensation temperature of the system as
well as the critical temperature. As it can be seen from Fig.\,8,
the compensation temperature of diluted system linearly decreases
for fixed $J$ and $D$ with increasing density of diluted sites in
the lattice, and on the other hand, compensation temperature of the
system systematically decreases when crystal field value is
increased. Finally, we can say that compensation temperature of the
diluted mixed system does not affected from lattice size as well as
the critical temperature.


In the present work, we have focused on the effect of non-magnetic
atoms on the critical and compensation temperatures of a two
dimensional mixed spin-1/2 and spin-1 system. Employing Monte Carlo
simulation method to the system, we have studied thermal and
magnetic behaviors of the system and dependence of these properties
on the non-magnetic atoms. We found that the thermal and magnetic
behaviors clearly depend on the number of non-magnetic atoms in the
lattice as it can be seen from the figures given. Particularly we
have shown that the critical and compensation temperatures linearly
decrease with increasing numbers of non-magnetic atoms of the two
dimensional mixed spin-1/2 and spin-1 system. Obtained results show
that dilution plays an significant role on the critical and
compensation points of a two dimensional mixed spin-1/2 and spin-1
system. On the other hand, these numerical results indicate that the
compensation temperature of the real ferrimagnetic spin systems can
be changed by the diluting the lattice with non-magnetic atoms, in
order to obtain desired compensation temperature.


\end{document}